\documentclass{article}
\usepackage{cite} 
\usepackage{spconf,amsmath,graphicx}
\usepackage{tabularx}
\usepackage[ruled,linesnumbered]{algorithm2e}
\usepackage[labelfont=bf]{caption} 
\usepackage{hyperref,url}
\usepackage{multirow}
\usepackage{booktabs}
\usepackage{amssymb}

\usepackage{spconf,amsmath,graphicx}

\title{Code-Switching Text Generation and Injection in Mandarin-English ASR}
\name{
\begin{tabular}{c}
Haibin Yu$^{1,2,\dag}$, Yuxuan Hu$^{2}$, Yao Qian$^2$, Ma Jin$^2$, \\
Linquan Liu$^2$, Shujie Liu$^2$,Yu Shi$^2$,  Yanmin Qian$^{1}$ ,Edward Lin$^2$,Michael Zeng$^2$
\end{tabular}
\thanks{$^{\dag}$Work is done by the first author during internship at Microsoft.}}

\address{
  $^1$MoE Key Lab of Artificial Intelligence, AI Institute\\X-LANCE Lab, Department of Computer Science and Engineering, Shanghai Jiao Tong University
  \\
  $^2$Microsoft Corporation\\
  {\small\{yuhaibin\_39, yanminqian\}@sjtu.edu.cn,\{yuxuanhu, yaoqian, majin1, linqul, shujliu, yushi, edlin, nzeng\}@microsoft.com}
  }
\begin{document}
\ninept
\maketitle
\begin{abstract}
Code-switching speech refers to a means of expression by mixing two or more languages within a single utterance. Automatic Speech Recognition (ASR) with End-to-End (E2E) modeling for such speech can be a challenging task due to the lack of data. In this study, we investigate text generation and injection for improving the performance of an industry commonly-used streaming model, Transformer-Transducer (T-T), in Mandarin-English code-switching speech recognition. We first propose a strategy to generate code-switching text data and then investigate injecting generated text into T-T model explicitly by Text-To-Speech (TTS) conversion or implicitly by tying speech and text latent spaces. Experimental results on the T-T model trained with a dataset containing 1,800 hours of real Mandarin-English code-switched speech show that our approaches to inject generated code-switching text significantly boost the performance of T-T models, i.e., 16\% relative Token-based Error Rate (TER) reduction averaged on three evaluation sets, and the approach of tying speech and text latent spaces is superior to that of TTS conversion on the evaluation set which contains more homogeneous data with the training set.  
\end{abstract}
\section{Introduction}
\label{sec:intro}
Code-switching (CS) is a linguistic phenomenon where different languages are alternated and spoken within the same utterance or context. Building a code-switching ASR system requires handling unpredictably switching languages within a single utterance. There can be some code-switching-specific variations on the boundary between the two languages, which are essential for ASR model to capture. However, the code-switching data is always insufficient to train a decent model. A considerable number of works have been proposed in acoustic feature extraction, model architecture design, and other prospects to improve code-switching ASR.

The less requirement of linguistic knowledge in building end-to-end (E2E) ASR has accelerated the  development of code-switching ASR recently. The most popular E2E model architecture like Connectionist Temporal Classification (CTC), attention-based sequence-to-sequence model, and transducer have been investigated for code-switching ASR \cite{luo2018towards,zeng19_interspeech,8683223,shan19investigate,9362075,chuang2021non,2022mwe}. In these works\cite{luo2018towards,zeng19_interspeech,shan19investigate}, the joint CTC-attention structure with language identification (LID) was constructed in a multi-task learning (MTL) framework. It exploits the advantages of both the attention and CTC while the additional LID module enables the system to handle switching languages. In the most recent works, multiple transformer-based structures have been employed to consider the property of code-switching. A bi-encoder transformer network-based Mixture of Experts (MoE) architecture has been proposed to capture language-specific information with a gating function performed as a language identifier\cite{lu20f_interspeech}. A multi-encoder-decoder (MED) Transformer architecture was used to capture the individual language attributes with multiple language-specific encoders and fuse them with multi-head attention in a decoder\cite{zhou20b_interspeech}. In addition, they pre-trained all of the language-specific modules by using large monolingual speech corpora. Self-supervised pretraining with multi-lingual unlabeled data without code-switching speech has also been shown effective to improve the performance of Mandarin-English code-switching ASR \cite{tseng2021mandarin}. 

Transducer-based ASR model is very attractive in the industry since it provides a natural way for streaming. However, it is not so straightforward to utilize unspoken text data except converting to synthesized speech or the features from TTS with a high computational cost\cite{zhao2021addressing,chen2021injecting}. There is no attention mechanism between the encoder and the decoder. Although the prediction module plays as a Language Model (LM), it is different from the conventional LM due to the blank token \cite{rao2017exploring} that makes it harder to leverage external LM. In this paper, we mainly focus on the Transformer-Transducer (T-T) architecture \cite{zhang2020transformer} and investigate cross-modality learning methods to leverage text-only data for improving the performance of the Mandarin-English code-switching ASR system. 
%


\section{Related work}
\label{sec:related}

 T-T models for code-switching ASR has not been intensively investigated so far. A recent work \cite{dalmia2021transformer} has demonstrated the improvements over the previous transducer model by leveraging several strategies including CTC and LM joint training, LID aware masked training, and multi-label/multi-audio encoder framework for additional monolingual corpora. The effectiveness of these strategies was presented only on a T-T model trained with a small-scale corpus named SEAME that contains about 100 hours of code-switching speech\cite{lyu2010seame}. The generalization capability of these strategies needs to be further exploited on an industry-scale corpus.
 
There are several works on unifying speech-text representation learning. Speech and text latent spaces were aligned via the shared layers from speech encoder and text encoder\cite{renduchintala18_interspeech,wangwei,bapna2021slam,tang-etal-2022-unified,chen22r_maestro, ao2021speecht5}, modality alignment loss \cite{bapna2021slam} like predicting whether a pair of speech and text is positive or negative, cross-modality learning loss, i.e., Mean Squared Error (MSE) between the embeddings from speech encoder and text encoder\cite{chen22r_maestro}, and cross-modality contrastive learning \cite{ye-etal-2022-cross}. These approaches can make speech representations learn more contextual information from text representations. To the best of our knowledge, these cross-modality learning methods have not been explored in the scenario of code-switching T-T based ASR. 


\section{Methods}
\label{sec:methods}
In this study, we build a Mandarin-English code-switching ASR system based on T-T structure. To address the lack of large-scale code-switching data in T-T modeling, we first generate code-switching text data and then leverage generated text into T-T modeling.

\subsection{Code-switching Text Data Generation}

\begin{figure}[t]
\includegraphics[width=0.8\linewidth]{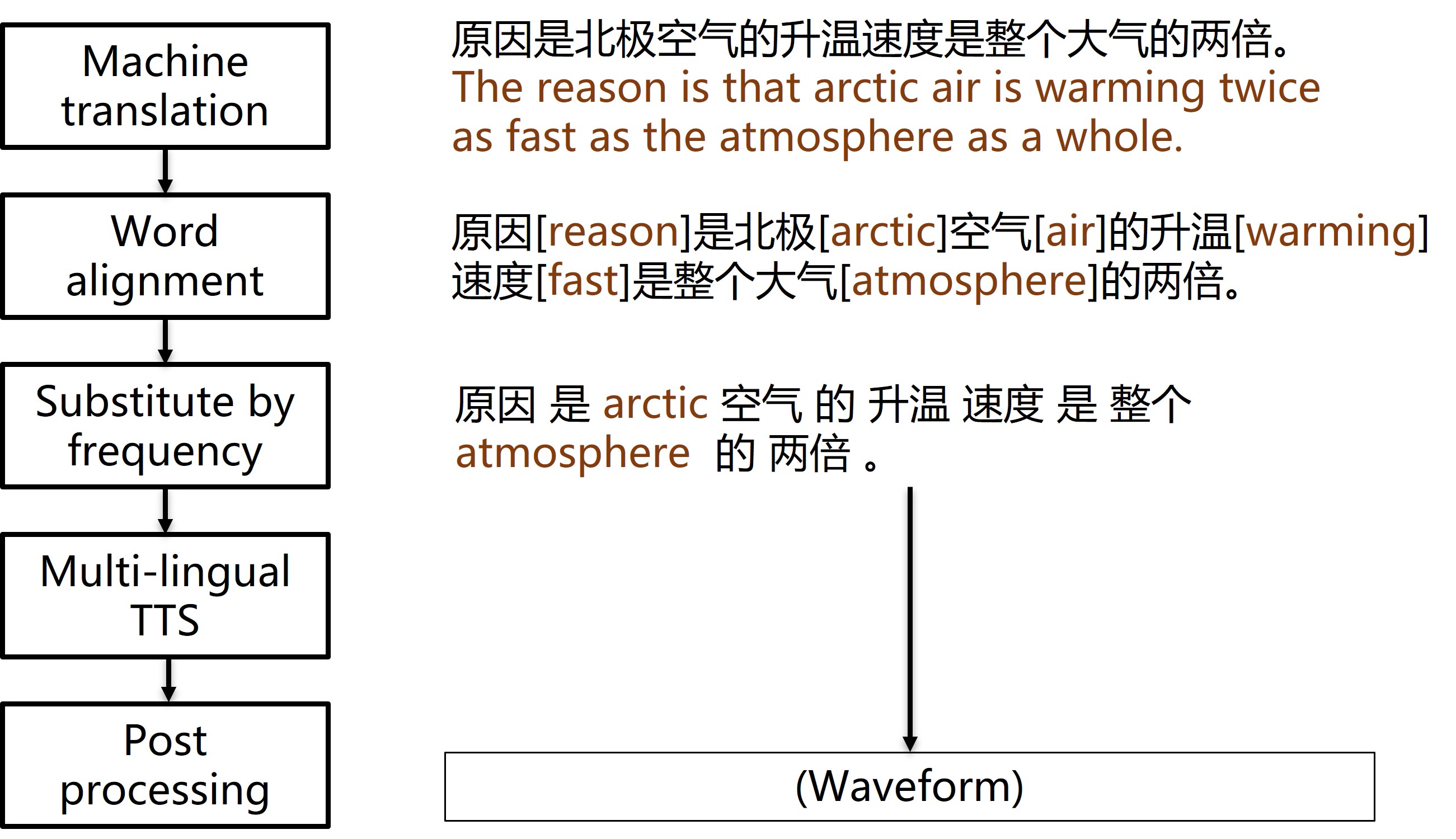}
\centering
\vspace{-1mm}
\caption{Code-switching data generation.}

\label{fig:TextGen}
\vspace{-5mm}
\end{figure}

The code-switching text data generation is illustrated in Fig. \ref{fig:TextGen}. First, parallel Mandarin and English sentence pairs are generated using a machine translation model. A word alignment procedure is then performed on the parallel sentence pairs to align the words with the same meanings but in different languages. Word alignment is conducted only on verbs and nouns. Finally, these verb and noun words are replaced across languages based on word frequency. English words occupy a percentage of about 10\% in the generated code-switching text data that meets the statistics of raw training data we collected in this study.

\begin{figure*}[ht]
\centering
\includegraphics[width=0.8\linewidth]{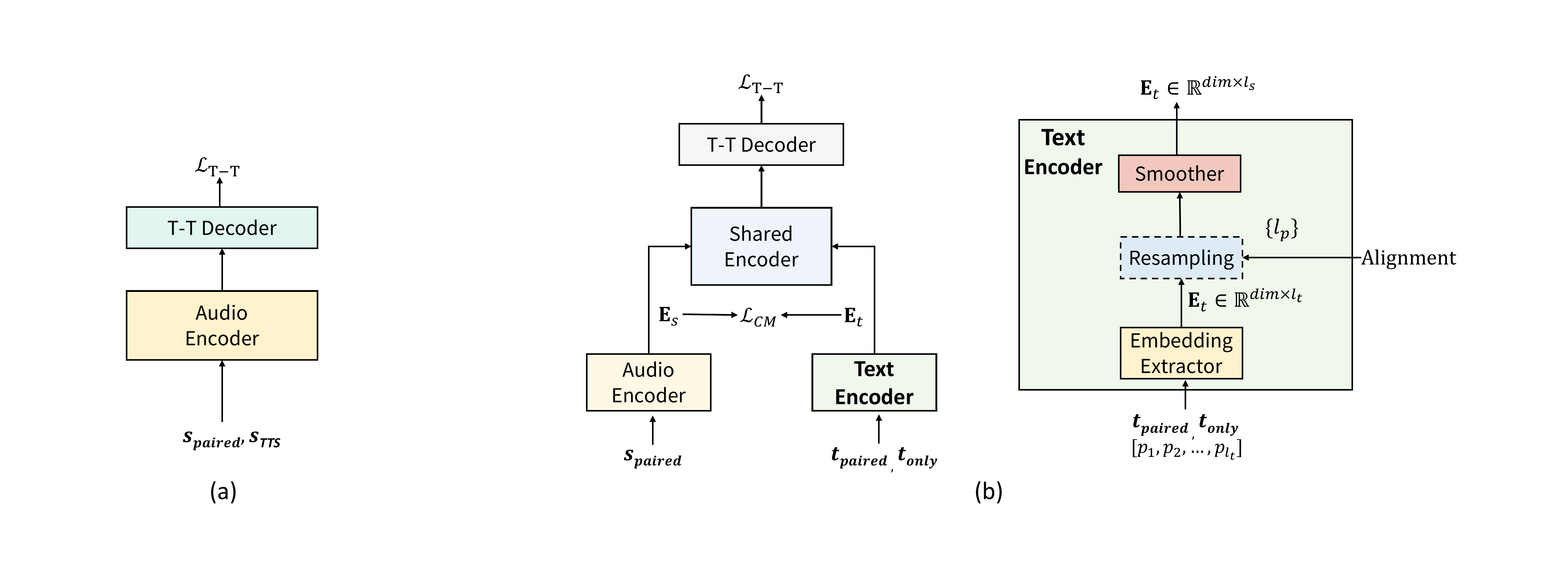}
\vspace{-8mm}
\caption{Code-switching text data injection by a) TTS conversion and b) cross-modality learning.}
\label{fig:CSASR}
\vspace{-5mm}
\end{figure*}

\subsection{Code-switching Text Data Injection}

We explore two approaches to inject code-switching text data into T-T modeling, as shown in Fig. \ref{fig:CSASR}, where illustrates a) augment paired speech-text training data by adding TTS converted data; and b) directly add text-only data by cross-modality learning. 

\subsubsection{Text-to-speech (TTS) Conversion}
We employ a multilingual TTS system to convert the generated code-switching text to speech. The system is based on Neural TTS that can mimic 18,000 different voices, so the generated tones of the speech are in a vast diversity. Text normalization is used in pre-processing, and potential low-quality audios are filtered out after generation. In addition, various reverberations and background noises are added to the speech to simulate different scenarios. The real paired speech-text training data, {$s_{paired}$}, is augmented with TTS converted data, {$s_{TTS}$}, to train a T-T model as shown in Fig. \ref{fig:CSASR} (a).

%
%
%

\subsubsection{Cross-modality Learning}

We investigate cross-modality learning to tie speech and text latent spaces and then inject text data into the tied space directly. As shown in Fig. \ref{fig:CSASR} (b), the text sequences can be derived from both paired speech-text data, {$t_{paired}$}, and text-only data, {$t_{only}$}.

The Speech Encoder is a stack of Transformer layers, which takes the speech sequence as its input and produce a speech representation $\mathbf{E}_s$. The text sequence is first converted to a phoneme sequence and then fed into the Embedding Extractor to get the embeddings. The embedding dimension is the same as that of speech representation $\mathbf{E}_s$. We denote the sequence length as $l_t$, and the embedding dimension as $dim$. The extracted text/phoneme embedding $\mathbf{Emb}_t$ belongs to $\mathbb{R}^{l_t\times dim}$. We upsample it to the length of $\mathbf{E}_s$ which has a much larger number of embeddings in terms of frames. The upsampler repeats the embeddings across the time domain according to the phoneme length $\{l_p\}$ provided by the force alignment result. The upsampled text/phoneme embeddings are afterward inputted into the Smoother, which is also a stack of Transformer layers, and finally the text representation $\mathbf{E}_t$ is calculated. Note that  $\mathbf{E}_t$ and $\mathbf{E}_s$ are of the same length. 

We align $\mathbf{E}_t$ and $\mathbf{E}_s$ into the same latent space by cross-modality learning loss. The details will be introduced in the next subsection. The Shared Encoder takes $\mathbf{E}_t$ and $\mathbf{E}_s$ separately as input, and the subsequent forwarding steps are similar to those in T-T. We apply a Transducer loss denoted as $\mathcal{L}_s$ for $\mathbf{E}_s$. Another Transducer loss $\mathcal{L}_t$ is applied to the decoding result from $\mathbf{E}_t$. The loss function is shown in equations \ref{eq:loss_paired1} and \ref{eq:loss_paired2}. In the following equations, the forward calculations of Speech Encoder, Embedding Extractor and Smoother are denoted as $\theta_s$, $\theta_{emb}$ and $\theta_{smooth}$ respectively, and the resampling process is denoted as $\mathrm{Resample}$. 
$\mu$ in the loss function is a hyperparameter to control the overall weight of the extra losses and the vanilla speech Transducer loss $\mathcal{L}_s$.
\begin{equation}
    \mathbf{E}_s=\theta_s(\mathbf{s}), \mathbf{Emb}_t=\theta_{emb}(\mathbf{t})
\end{equation}
\begin{equation}
    \mathbf{E}_t=\theta_{smooth}(\mathrm{Resample}(\mathbf{Emb}_t, \{l_p\}))
\end{equation}
\begin{equation}
    \mathcal{L}_{paired}=\mu\mathcal{L}_{s} + (\mathcal{L}_{CM} + \mathcal{L}_t)
   \label{eq:loss_paired1}
\end{equation}
\begin{equation}
   \mathcal{L}_{s} = \mathcal{L}_{\mathrm{T-T}}(\mathbf{E}_s, \mathbf{t}), {\mathcal{L}_{t} = \mathcal{L}_{\mathrm{T-T}}(\mathbf{E}_t, \mathbf{t})}
\label{eq:loss_paired2}
\end{equation}

Two cross-modality loss functions, $\mathcal{L}_{CM}$, and one learning strategy are explored in this study. 
\begin{enumerate}
    \item MSE 
    
     MSE loss is a strict criterion that forces the speech representation $e_s$ and text representation $e_t$ to be identical. 

    \item BiInfoNCE 
    
     Apart from MSE loss, other soft criteria are worth testing, like InfoNCE loss that lifts the lower bound of the mutual information of the two representations\cite{oord2018representation}. Here we employ BiInfoNCE loss illustrated as equation \ref{equa:biinfonce} and \ref{equa:infonce}, where $f$ is cosine similarity.
     \begin{equation}
        \mathcal{L}= \mathcal{L}_{\mathrm{N}}(\mathbf{E}_t, \mathbf{E}_s) + \mathcal{L}_{\mathrm{N}}(\mathbf{E}_s, \mathbf{E}_t)
        \label{equa:biinfonce}
     \end{equation}
     \begin{equation}
        \mathcal{L}_{\mathrm{N}}(X,Y)=-\underset{X}{\mathbb{E}}\left[\log \frac{f\left(x_{i+k}, y_{i}\right)}{\sum_{x_{j} \in X} f\left(x_{j}, y_{i}\right)}\right],y_{i} \in Y\\
        \label{equa:infonce}
     \end{equation}
     
    \item Modality swap
    
     In this approach, no cross-modality loss is employed. Instead, by swapping samples from $e_s$ and $e_t$, during the optimization procedure, the latent space of $e_s$ and $e_t$ could be tied together.
\end{enumerate}

For the generated text-only input, {$t_{only}$}, similar to the text modality, {$t_{paired}$}, in the paired speech and text input, the Text Encoder extracts embedding $emb_t$ from the text sequence, $emb_t$ is upsampled and smoothed into text representation $e_t$, and the Shared Encoder takes $e_t$ as input and the following modules complete the Transducer computation. The backward process traverses the decoder, shared encoder, and text encoder, which means Speech Encoder will not be optimized. The loss function for text-only input is the Transducer loss $\mathcal{L}_t$ applied to the decoding result from $e_t$, as is represented in equation \ref{eq:loss_tonly}.
\begin{equation}
   \mathcal{L}_{text} = \mathcal{L}_{t} = \mathcal{L}_{\mathrm{T-T}}(\mathbf{E}_t, t)
    \label{eq:loss_tonly}
\end{equation}

\section{Experiments}
\label{sec:exp}

\subsection{Data and Systems}

We use a corpus containing 1.8k (one thousand eight hundred) hours of real intra-sentential code-switching data as paired speech and text input, already augmented by adding noise and speed perturbation. For the generated text-only input, to line up the acoustic and the linguistic information, we use the phoneme sequences and upsample them according to the duration model of the TTS system. The TTS-generated code-switching corpus contains 14k hours of speech.
Three systems: 1) baseline: T-T was trained with 1.8k hours of real speech; 2) topline: T-T was trained with real and TTS converted speech as shown in Fig.\ref{fig:CSASR} (a); 3) CM: cross-modality learning is used in T-T modeling as shown in Fig.\ref{fig:CSASR} (b), are built and tested on three internal code-switching evaluation sets in the scenarios of online meeting, language teaching and general dictation with a total duration of 60 hours. The third set is homogeneous to part of the training data and the rest two sets consist of out-of-domain data.  

\subsection{Model Configuration and Training}

For the baseline and topline systems, we use a plain Transformer-Transducer configuration similar to the original implementation, which has 18 layers in the audio encoder and 2 layers in the decoder. For the cross-modality learning systems, to keep consistency, the Speech Encoder contains 6 Transformer layers, and the Shared Encoder has 12 Transformer layers. The Embedding Extractor is a stack of 4 Transformer layers, and the Smoother has 2 Transformer layers. The input feature or embedding size is 512. Each attention sub-layer has 8 heads without dropout, and the dropout ratio of all other layers is 0.1. The learning rate schedule is warmed up to $2$e$-4$ in the first 50k steps and then decays exponentially. We train all models using AdamW optimizer. The text encoder is only involved in the training stage and discarded during inference. 

In the cross-modality (CM) learning systems, paired speech-text and text-only inputs are not mixed in a single batch. Instead, each batch is composed of two mini-batches, one of which contains the paired speech and text samples, and the other consists of text-only samples. The gradients are accumulated within one single batch. Every time the model finishes two mini-batches in one batch, the gradients are back-propagated and the model parameters are updated.


\subsection{Metrics}

We measure the performance of our systems in terms of the overall token error rate (TER). In order to have a detailed analysis, we also calculate the word error rate (WER) on the English part and the character error rate (CER) on the Mandarin part of the code-switching data, respectively. 

\section{Results}

\subsection{Cross-modality Learning and Generated Text}

The overall results achieved by baseline, topline, and CM systems are reported in Table 1.

\begin{table}[htb]
\label{tab:overall}
  \centering
  \begin{tabular}{cccc}
    \toprule
    \textbf{System} &CS-All &CS-Man &CS-Eng  \\
    \midrule
    baseline            &15.6   &11.05  &40.05\\
    CM MSE, $\mu=2$      &14.2   &10.06   &36.78\\
    CM MSE, $\mu=2.33$      &13.2   &9.13   &35.4\\
    CM MSE, $\mu=3$      &15.0   &10.92   &37.35\\
    CM MSE, $\mu=5$      &15.4   &11.2   &37.91\\
    topline             &13.1   &10.23  &28.39\\
    
    \bottomrule
  \end{tabular}
  \caption{Overall TER\%, CER\% and WER\% on three code-switching evaluation sets. CS-Man and CS-Eng represent the Mandarin and English parts, respectively.}
\vspace{-5pt}
\end{table}

\begin{table*}[ht]
\label{tab:breakdown1}
  \centering
  \begin{tabular}{cccc|ccc|ccc}
    \toprule
    \multirow{2}{*}{\textbf{System}} &\multicolumn{3}{c|}{Set 1} &\multicolumn{3}{c|}{Set 2} &\multicolumn{3}{c}{Set 3} \\
                                    &CS-All &CS-Man &CS-Eng    &CS-All &CS-Man &CS-Eng    &CS-All &CS-Man &CS-Eng\\
    \midrule
    baseline                        &13.22	&12.54 &15.81 &20.78 &12.8 &66.78 &5.91 &4.36 &21.62\\
    CM MSE, $\mu=2.33$    &11.52 &10.68 &14.68 &17.21 &10.18 &57.72 &5.71 &4.11 &21.99\\
    topline             &12.1 &12.15 &11.93	&15.86 &10.5 &46.8 &7.33 &6.6 &14.7\\
    
    \bottomrule
  \end{tabular}
  \caption{A breakdown of TER\%, CER\% and WER\% on three code-switching test sets from baseline, topline and cross-modality learning systems. Set 3 is the homogeneous set.}
\vspace{-2mm}
\end{table*}

\renewcommand\thetable{4}
\begin{table*}[ht]
\label{tab:breakdown2}
  \centering
  \begin{tabular}{cccc|ccc|ccc}
    \toprule
    \multirow{2}{*}{\textbf{System}} &\multicolumn{3}{c|}{Set 1} &\multicolumn{3}{c|}{Set 2} &\multicolumn{3}{c}{Set 3} \\
                                    &CS-All &CS-Man &CS-Eng    &CS-All &CS-Man &CS-Eng    &CS-All &CS-Man &CS-Eng\\
    \midrule
    CM MSE, $\mu=2.33$    &11.52 &10.68 &14.68 &17.21 &10.18 &57.72 &5.71 &4.11 &21.99\\
    Swap rate 0.2             &12.77 &12.06 &15.45	&20.25 &12.68 &63.93 &4.83 &3.3 &20.34\\
    BiInfoNCE                       &12.67	&17.87 &15.72 &20.05 &12.75 &62.12 &4.96 &3.46 &20.2\\
    
    \bottomrule
  \end{tabular}
  \caption{A breakdown of TER\%, CER\% and WER\% on three code-switching test sets by using different cross-modality learning mechanics.}
\vspace{-8pt}
\end{table*}

Table 1 shows that adding generated text or synthesized speech into the training procedure significantly boosts the performance of the T-T over the baseline by a relative 16\% of WER reduction (when $\mu=2.33$). The overall performance of the cross-modality learning model is on a par with that of the topline model. It demonstrates that generating code-switching data is a promising approach to improve the performance of T-T based code-switching ASR. Furthermore, the synthesized speech is different from natural CS speech and may contain TTS-specific artifacts that hurts the model performance on some testing samples. Cross-modality learning approaches only utilize the generated text without actual speech conversion and may have a chance to reduce the impacts of artifacts. As English words occupy a percentage of about 10\%, the code-switching boundary patterns are vital to its recognition performance. The lower English WER compared to baseline suggests that the code-switching boundary patterns can be captured in the generated text by using cross-modality methods. We attempted to remove $\mathcal{L}_{CM}$ in equation \ref{eq:loss_paired1}. The resultant performance shows only a marginal gain over the baseline system. 

The breakdown results of the three sets are given in Table 2, in which it shows that the cross-modality learning system can outperform both the baseline and topline systems on the homogeneous set.

It is noteworthy that all the systems report much higher WER on the English part of CS speech compared with the CER on the Mandarin part. The inconsistent performance in different languages results from the imbalance of English and Mandarin in both training and test data sets. As mentioned in 3.1, English words occupy a small proportion of the CS speech and are difficult for the model to recall. In some typical error cases, the English words in the utterances are misrecognized as Chinese characters with identical or similar pronunciation. These results reflect that detecting the boundary between the languages is still a hard task.

The improvement on the Mandarin part is more responsible for the overall performance gain of the CM systems. Comparing the CM system with the baseline, the benefits from combining data of extra modality into training are mainly seen in Mandarin. The topline system is inferior to the CM systems in the aspect of Mandarin part but reports the lowest WER of English part. This can be interpreted as the other side of the coin. The cross-modality learning procedure takes more advantage of the density of Mandarin, in contrast to the sparsity of English. Since Mandarin takes up the majority of the generated text and the real code-switching speech, if only the Mandarin parts from the two modalities were aligned, a large part of the cross-modality learning would be well conducted. In other words, cross-modality learning is more focused on the Mandarin part, but may ignore the uncommon patterns to some extent. This can lead to a less optimal result for the model.

\subsection{Other Cross-modality Learning Mechanics}

The results of cross-modality learning approaches other than MSE are shown in Tables 3  and 4. For the BiInfoNCE loss, $\mu$ is set to 2.33. For the modality swap strategy, the swap rate is set to 0.2, which means a random 20\% of the frames from $e_t$ are substituted for the corresponding frames from $e_s$.

\renewcommand\thetable{3}
\begin{table}[ht]
\label{tab:multimethod}
  \centering
  \begin{tabular}{cccc}
    \toprule
    \textbf{System} &CS-All &CS-Man &CS-Eng  \\
    \midrule
    baseline            &15.6   &11.05  &40.05\\
    CM MSE, $\mu=2.33$      &13.2   &9.13   &35.4\\
    Swap rate 0.2 &15.0	&10.63	&38.44\\
    BiInfoNCE &14.89	&10.65	&37.7\\
    \bottomrule
  \end{tabular}
  \caption{Overall TER\%, CER\% and WER\% by using different cross-modality learning mechanics.}
\vspace{-4mm}
\end{table}

From the results, although all variations of cross-modality learning systems can outperform the baseline, the system using MSE loss still takes the leading place. The effectiveness of generated text data is validated by these results acquired with multiple cross-modality learning approaches. Besides, it also suggests that the generated text data alone cannot guarantee an expected performance boost, and intentionally aligning the representations from the two modalities into the same subspace  is vital to the performance. We observe that two CM systems of BiInfoNCE and modality swap reach a lower error rate than that of MSE on the homogeneous test set shown in Table 4.  
\subsection{Preliminary Studies on Other Data Augmentation}
We tried to leverage multi-lingual unlabeled speech to improve the performance of our code-switching model as \cite{tseng2021mandarin}. The experiment carried out on the model pre-trained in a self-supervised manner with 75k hours of speech in 10 languages shows a small performance gain over the model trained with 1.8k CS real data only but the gain disappears when the generated CS data is added. In addition, we added mono-lingual Mandarin and English labeled data (each of them contains 10k hours of speech) into our model training together with an auxiliary task of language identification at the frame level. It can significantly improve the model performance on mono-lingual evaluation sets. However, the performance improvement over the model trained with both CS real and generated data is marginal on the code-switching test sets, which we are concerned about in this work.       

\section{Conclusions}

In this paper, we demonstrate that it is beneficial to utilize external data via generating code-switching text and using cross-modality learning methods to inject it. We also investigate three approaches of cross-modality learning to tie the speech and text representations closer. It helps the T-T model to capture more code-switching-specific features and thus boosts the performance of code-switching Mandarin-English ASR. In future works, we will explore various cross-modality learning more deeply and seek for other approaches to leveraging external data.

\vfill

\newpage
 \let\oldthebibliography\thebibliography
 \let\endoldthebibliography\endthebibliography
 \renewenvironment{thebibliography}[1]{
 \begin{oldthebibliography}{#1}
 \setlength{\itemsep}{0.1em}
 \setlength{\parskip}{0.1em}
 }
 {
 \end{oldthebibliography}
 }
\bibliographystyle{IEEEbib}
\bibliography{mybib}

\end{document}